# Temperature-dependent electron microscopy study of Au thin films on Si (100) with and without native oxide layer as barrier at the interface

A Rath[1], J K Dash[1], R R Juluri[1], A Rosenauer[2] and P V Satyam[1,2,*,$]

[1] Institute of Physics, Sachivalaya Marg, Bhubaneswar - 751005, India

[2] Institute of Solid State Physics, University of Bremen, D-28359 Bremen, Germany

**Abstract**

Real time electron microscopy observation on morphological changes in gold nanostructures deposited on Si (100) surfaces as a function of annealing temperatures has been reported. Two types of interfaces with the substrate silicon were used prior to gold thin film deposition: (i) *without* native oxide and on ultra-clean reconstructed Si surfaces and (ii) *with* native oxide covered Si surfaces. For a ≈ 2.0 nm thick Au films deposited on reconstructed Si (100) surfaces using molecular beam epitaxy method under ultra high vacuum conditions, aligned four-fold symmetric nanogold silicide structures formed at relatively lower temperatures (compared with the one with native oxide at the interface). For this system, 82% of the nanostructures were found to be nano rectangles like structures with an average length ≈ 27 nm and aspect ratio of 1.13 at ≈ 700°C. For ≈ 5.0 nm thick Au films deposited on Si (100) surface with native oxide at the interface, formation of rectangular structures were observed at higher temperatures (≈ 850° C). At these high temperatures, desorption of the gold silicide followed the symmetry of the substrate. Native oxide at the interface was found to act like a barrier for the inter-diffusion phenomena. Structural characterization was carried out using advanced electron microscopy methods.

[*] Corresponding Author: satyam@iopb.res.in, pvsatyam22@gmail.com

[$] On leave from Institute of Physics, Bhubaneswar – 751005, India

# 1. Introduction

As the dimensions of electronic devices decrease with increasing packing density, the thickness of metal layers decreases continuously. An understanding of the growth of metal thin film on a semiconductor becomes important both from fundamental and practical points of view. The interaction of metal films with silicon has attracted considerable attention because of its importance in semiconductor technology. Metal nanoparticles on surfaces are being used effectively in nanofabrication, as catalysts for carbon nanotube growth [1, 2], catalysts for the vapor-solid-liquid (VLS) growth of inorganic nanowires [3 – 5], and for nano electronics applications [6 – 8]. Formation of nano–metal silicides, used as Schottky barriers and ohmic contacts in devices, is of particular interest in the emerging area of nanoscience and nanotechnology. The mechanism of these interactions, the range of temperatures required and the resulting composition of Au-Si alloy depends on whether these gold nano islands interact directly with the substrate or they will have to deal with a barrier like native oxide layer before, they could interact with substrate. Either of the processes may give rise to various kinds of self assembled nano/micro structures depending on the symmetry of the substrate. The role of nanosilicides is important in many ways and hence a deeper understanding at atomic scale for the silicon/metal interfaces is indispensable [9]. Recently, fabrication of micro/nanoscale pits on Si substrates with long range ordering and reliable shape control *via* a facile vapor transport method assisted by Au nanoparticles has been achieved [10]. The resulted pits in the shape of triangles, squares and, wires were obtained on Si (111), (100) and (110) substrates, respectively [10]. How the desorption takes place to form regular shaped pits would be interesting to study. This paper addresses a part of this problem as we focus on the desorption of gold at high temperatures from the regular shaped gold silicide structures.

Self-assembly is an attractive nanofabrication technique because it provides the means to precisely engineer the structures on the nanometer scale over large sample areas. Self-organizing nano crystal assemblies have already shown the degree of control necessary to address the challenges of building nanometer-scale technologies [11]. A systematic study of such processes has implications on modern day semiconductor technology. Studies of processes going on at the Si/ metal interfaces at low temperatures have been well attempted by several groups [11 – 20]. Most of these studies were done under UHV conditions and on systems where there was no oxide layer in between the metal film and the silicon substrate. Many of them have reported the formation of various metastable gold silicides due to the inter mixing of silicon and metal. There have been numerous interesting studies on the mechanism of relaxation of strain in such epitaxial layers. Strained epitaxial layers are inherently unstable and have interesting properties. It has been reported that the dislocation formation and the shape transition are the important processes by which relaxation of strain occurs [21 – 24]. In recent years, it has been recognized that shape changes such as island formation constitute a major mechanism for strain relief [21, 23]. Tersoff and Tromp reported that a strain-induced shape transition may occur [21]. Below a critical size, islands have a compact symmetric shape. For larger sizes, they adopt a long thin shape that allows better elastic relaxation of the island's stress [21].

Metal silicide thin films on Si(100) have been studied in great detail for more than two decades due to the possibility of obtaining self-aligned epitaxial metal-semiconductor interfaces [19,22]. Sekar et al. [23, 24] reported the formation of silicide when a 100 nm Au film deposited on Br-passivated Si(111) was thermally annealed. The annealing treatment of the as-deposited Au/Si at eutectic temperature (~363°C) gave rise to silicide of composition close to $Au_4Si$ as inferred from Rutherford backscattering spectrometry (RBS) studies. Mundschau et al. reported the formation of triangular islands, when sub monolayer gold was deposited epitaxially on Si(111) substrate under UHV conditions [25]. Vacuum annealing of thick and continuous gold films deposited on passivated Si(110) surface showed the formation of high aspect ratio gold silicides ($Au_4Si$) wire like structures around 363 °C [26]. Earlier, we showed the formation of well-aligned low aspect ratio gold silicide nanostructures that were grown under UHV conditions at temperatures in the range 600–700 °C in the absence of native oxide at the interface [27]. Using real time *in situ* high temperature transmission electron microscopy (with 40 ms time resolution), we showed the formation of high aspect ratio (≈15.0) aligned gold silicide nanorods in the presence of native oxide at the interface during *in situ* annealing of gold thin films on Si(110) substrates [28]. This was explained using an oxide mediated liquid–solid growth mechanism. For other substrate surfaces, such as Ni(100) substrate, homo- and hetero-epitaxial growth has been studied, in which E. Kopatzki *et al* have reported the formation of nickel nano squares /rectangles of average size 2 nm [29]. Similarly, Frank *et al* have reported the formation of rectangles /squares in the homo epitaxial system of Ag on Ag (100) substrate [30]. For gold on Si (100) substrates, a detailed study on the preferred orientation growth of gold under $N_2$ atmospheric conditions was carried out and found that structures reorient themselves at higher temperatures [31]. In this paper, no details on possible silicide formation and desorption has been reported. The knowledge of real time shape variations, and structural phase transitions in Au/Si system would help to properly understand silicide growth at nano scale and would form a basis for appropriate applications that are based on the use of gold silicide islands as catalysts. As the Si (100) substrate is used for many electronic applications in semiconductor industry, it will be very useful to understand nucleation, growth of gold silicide nanostructures and desorption as a function of temperature.

In this paper, we present the experimental observation of the growth of aligned gold silicide nano-structures during *in-situ* thermal treatment in a TEM heating stage for gold deposited on Si (100) surfaces with and without an oxide layer at the interface. Followed by this, *ex-situ* STEM measurements and high resolution electron microscope measurements confirm the presence of gold in these rectangular shaped silicide structures and also show depletion of gold due to desorption of gold silicide at high temperatures.

## 2. Experimental

A thin Au film of thickness ≈2 nm was deposited on n-type Si (100) of resistivity 10–20 Ω cm, by the molecular beam epitaxy (MBE) method under UHV (base pressure ≈ 2 × $10^{-10}$ mbar) conditions [32]. The Si(100) substrates were loaded into the MBE chamber and degassed at ≈ 600°C for about 12 hours inside the chamber, followed by flashing for about 3 minutes by direct heating at a temperature of ≈1200°C. In this process, native oxide was removed and a clean Si (100) surface was obtained. On such ultra clean surfaces, ≈ 2.0 nm thick gold films were grown epitaxially by evaporating Au from a Knudshen cell. Deposition rate was kept constant at ≈ 0.14 nm $min^{-1}$. During the growth, the chamber vacuum was ≈ 6.2 ×$10^{-10}$ mbar. The thickness monitor was calibrated with RBS measurements. After the deposition, the sample was taken out of MBE chamber. We have also deposited Au films of about ≈ 5.0 nm thickness on n-type Si (100), by the thermal evaporation method under high vacuum (≈ 4 × $10^{-6}$ mbar) conditions. For this case, native oxide was kept intact and the thickness was monitored using a quartz crystal monitor. Planar TEM specimens were prepared from the above samples. Disks of ≈ 3 mm diameter were cut using an ultrasonic disc-cutter followed by lapping until it reached ≈ 100 μm thickness. Using a dimple grinder, samples were further thinned down to ≈ 25 μm at the center. Finally, electron transparency was achieved at low energy $Ar^+$ ion milling. Temperature dependent TEM measurements were done with 200 keV electrons (2010, JEOL HRTEM) at Institute of Physics, Bhubaneswar (IOPB), India. A single tilt heating holder (GATAN 628UHR, that has a tantalum furnace and can heat the specimen up to 1000°C) was used. The temperature is measured by a Pt/Pt-Rh thermocouple and is accurate within a couple of degrees. The holder has a water cooling system to avoid over heating of the sample surroundings and the specimen chamber, while keeping only the sample at a specified temperature. Real time measurements were carried out using a CCD camera (GATAN 832) in which real time movies can also be recorded. For the ≈ 5 nm Au on Si(100) system, temperature dependent TEM measurements (*in-situ*, real time) were carried out using 200 keV system (TEM at IOPB) and *ex-situ*, room temperature (RT) and high resolution measurements were carried out using STEM and high resolution TEM (HRTEM) measurements using 300 keV electrons in the $C_s$-corrected FEI Titan 80/300 system at the University of Bremen, Germany.

## 3. Results and discussions

In-situ heating experiments were carried out at a ramp rate of 7°C min$^{-1}$. At each set of temperatures, the specimen was kept for about 30 minutes to achieve a stable temperature. We discuss about the 2 nm Au on Si (100) system without the native oxide barrier at the interface. As shown in Fig. 1(a), the gold islands have irregular shaped isolated structures in *as-deposited* system. The selected area diffraction (SAD) pattern shown in Fig. 1(b) confirms the polycrystalline nature of gold films. A close look at the diffraction pattern suggests the existence of some texturing, as more reflections from {220} plane can be observed. A high resolution TEM taken on nanostructures shows lattice spacing corresponding to elemental gold (Figure 1C) from one of the isolated island structures. Figure 2 depicts bright field (BF) images at various temperatures (200ºC, 300ºC, 325ºC, 350ºC, 363ºC, 400ºC, 510ºC, 600ºC and 700ºC) and Figure 3 shows SAD patterns taken at various temperatures (325°C, 350°C, 363°C, 600°C and 700°C). At 200°C, as shown in Figure 2(a), an almost similar morphology was found as in the as-deposited case. This shows that the system was quite stable without any inter-diffusion among the gold nanostructures below 300°C (Fig. 2 (a) and (b)). Fig. 2(c) depicts a BF image taken after the system was kept stable at 325°C. Interestingly, the shape of nanostructure appears to start having definite symmetries (like 4- fold) at these low temperatures. During the observation, we noted that some rectangular structures started forming at many places. The real time SAD pattern at 325°C (i.e., while the system was kept at this temperature) is shown in Figure 3(a). The SAD pattern confirms the presence of silicon and un-reacted gold in the system. This shows that silicide formation is not a necessary condition to have well defined and symmetric structures. In-plane diffusion depending on the substrate orientation appears to play a major role in various shape formations of these gold nanostructures. It is well known that *bulk* Au-Si system has a eutectic phase at 363°C. The temperature was increased by smaller intervals (< 25°C) around the *bulk* eutectic temperature. Fig. 2(d) depicts a BF image taken at 350°C. At this temperature, a larger number of nano structures having rectangular shapes were formed. As the temperature was increased to 363°C and then to 400°C (Fig. 2(e), 2(f)), the coverage area of rectangular shaped nano structures increased. Also some of the structures were of square shape. The SAD pattern taken at 363°C (Fig. 3 (c)) showed signature (compared with the case at 325°C) of un-reacted gold at this temperature. We did not take SAD pattern after cooling the specimen to room temperature (RT) at this stage and hence it is not possible to comment on eutectic phase formation. It is important to note the studies by Ijima *et al* on the dynamic behaviour of ultra-fine gold particles (~2 nm) at the atomic resolution using a HRTEM equipped with real time video facility [33]. They observed that particles continually changed their shapes, orientations and internal atomic arrangements within a time interval of one tenth of a second. The in-situ diffraction patterns were found to be diffused, giving an impression that the particles are in liquid phase. But, after cooling down to room temperatures, high resolution TEM showed a regular arrangement of atoms in individual particles [33]. In our studies, at lower temperatures (< 363°C), gold nanostructures were relatively stable at these temperatures. Cooling to RT could help us in a proper determination of silicide formation at these low temperatures (< 363°C).

Real-time *in-situ* heating measurements were also carried out at 400°C (BF image at Fig. 2(f)), 510°C (Fig. 2(g)), 600°C (Fig. 2(h)), and 700°C (Fig. 2(i)). More aligned nano-rectangles/squares were formed at these higher temperatures. The average ratio of length to breadth, known as "aspect ratio", has been determined by using many bright field images taken at each temperature: 325, 350, 363, 400, 510, 600 and 700°C. In determining the aspect ratios, the longer side of the rectangles was taken as length. The average length, average aspect ratio and the percentage of nanostructures having square /rectangle shapes at various temperatures are tabulated in Table 1. At 325°C, the aspect ratio was found to be $1.30 \pm 0.02$ and the average length was about $19.5 \pm 1.1$ nm. Interestingly, even at this stage, about 36% of the nano - structures were having rectangular shape. As the temperature was increased, the average length and percentage of nano rectangles increased but the average aspect ratio decreased. At 363°C, the average aspect ratio and length was found to be $1.18 \pm 0.01$ and $24.3 \pm 1.0$ nm. At 400°C, average length was increased slightly and the aspect ratio was almost same (as at 363°C). This indicated that there was a small change in one of the side. The percentage of nano rectangles was found to be 69% at this temperature. After 510°C onwards, the average length, average aspect ratio and percentage of particles having rectangle /square shape were almost constant. At 700°C, the average aspect ratio and length was found to be $1.13 \pm 0.02$ and $25.9 \pm 1.3$ nm, respectively, and the percentage of particles having rectangle/square shape was 82%. This is an interesting point of observation. The aspect ratio larger than 1.0, is attributed to some kind of shape transformation. Such shape transformation occurs as a part of release of strain at the interfaces or within the island structures. During the silicide growth at high temperatures, higher aspect ratio silicide could be formed so as to reduce the strain energy, which is often explained by coincidence site lattice matching mechanism. In the present case, the variation in the aspect ratio is about 10 – 20 % over the temperature range, and cannot be attributed to the coincidence site lattice matching mechanism. From Fig.2, one can also observe that initially the gold nanostructures were having very rough edges and corners. With the increase in temperatures, smoothening of these edges and corners was observed. Ressel *et al* have reported about the behaviour of liquid Au-Si alloys on Si surfaces when ≈ 2.0 nm of gold was deposited on Si(111) and Si(100) substrates at 400°C under UHV condition to form the droplets [34]. Detailed and elaborate real time SEM measurements by this group discussed many aspects of wetting phenomena of Au – Si liquid alloys but as the characterization was done using SEM, crystalline quality details of these structures were not discussed [34]. Shape of the droplets changed from circular to hexagonal above 750°C for the Au/Si(111). But, in the case of Au/Si(100) droplets were octagonal when solid and on melting became rounded at higher temperatures. The authors attributed this to the anisotropy in the line tension between solid/liquid/gas lines. With increasing temperature, the entropic effects reduce the free energy difference between straight and curved steps leading to a rounding of the shape. With increasing temperature the entropic effects reduce the free energy difference between straight and curved steps leading to a rounding of the shape [34].

In the following, we discuss the SAD and high resolution transmission electron microscopy (HRTEM) results that are obtained for a system at different temperatures and at room temperature (after cooling from 700°C). Real time SAD patterns showed the diffraction spots of single crystalline silicon and rings of gold. Apart from this, a few extra spots and ring were also found which match neither with silicon nor gold. Interestingly, from 325°C to 350°C, no extra spots other than pure silicon spot and gold ring were seen for real time SAD (Fig. 3(a)-3(b)). After 363°C onwards, diffuse ring of d-spacing 0.235nm were observed along with silicon spots (Fig. 3(c)). This d-spacing value (0.235nm) matches with both gold and gold silicide phases. At $510^0$C, the SAD pattern shows the single crystalline spots of silicon and the diffuse ring of d-spacing 0.235nm (Fig. 3(d)). More careful analysis reveals that the edges of the nano rectangles are aligned along <110> and <1$\bar{1}$0> direction with respect to the silicon substrate. At 600°C, few extra reflections other than silicon and gold spots were observed (Fig. 3(e)) which were found to match with at least three reported phases of gold silicides. The spot nos.1, 2, 3 and 4 corresponds to a d-spacing of 0.253 nm, 0.453 nm, 0.356 nm and 0.469 nm respectively. When compared with the reported phases these lattice spacing match with several phases like $Au_5Si_2$, $Au_5Si$, $Au_7Si$, $Au_2Si$ and $Au_3Si$. Similarly at 700°C, spot nos. 1 and 2 and ring no 3 corresponded to a d-spacing of about 0.147 nm, 0.261 nm and 0.294 nm respectively (Fig. 3(f)). They matched with the phases of $Au_5Si_2$, and $Au_5Si$.

Figure 4(b) gives the SAD pattern obtained at room temperature (RT) after cooling down the sample from 700ºC. Apart from Si and Au reflections we found a few extra rings which were found to match with at least four reported phases of gold silicide. Ring no. 1, 2 and 3 corresponds to a d-spacing of about 0.133 nm, 0.156 nm and 0.261 nm respectively. On comparing them with the reported phases, these spacing match with several phases like $Au_5Si$, $Au_3Si$, $Au_7Si$ and $Au_5Si_2$. .Many diffraction patterns of gold silicides have been presented over the years [35 - 40]. Hultman *et al*. rightly pointed out that when comparing the observed data with reported phases of gold silicides, surprisingly there is a good agreement between the *d* values for at least 8-12 strongest reflections of all the alloys [41]. Looking at the constancy of lattice spacing of strongest lines, they doubt whether these diffraction patterns arise from different phases or they represent various degrees of ordering of super lattices of one fundamental structure. Because, the fundamental lines are similar for all reported phases except $Au_5Si_2$. As in our case, lack of symmetry further adds to the difficulty of assigning one definite phase to the obtained diffraction pattern. It should be noted that the minimum size of the SAD (SAD aperture) available with our system is bigger than a single silicide/island. Hence the SAD pattern shows these features (representing various reflections). Figure 4(a) and 4(c) shows the bright field image of the nano structures and the high resolution image of a single nano rectangle. From the bright field image one can observe that some moiré fringes are present on the nano rectangles and that the corners of the rectangles are rounded. Due to the different bonding strengths along two facets the migration barrier experienced by the gold atoms along the length of the rectangle and around the corners differ, which causes the rounding of the corners [42]. In summary, for 2 nm Au on clean Si (100) surfaces, ordered and symmetric structures were formed at higher temperature. The diffraction data suggest formation of nano silicide structures at higher temperatures.

We now present the experimental results on second case where a thin native oxide layer is present at the interface of Si (100) substrate and 5 nm Au film. Figure 5(a) and 5(b) show the bright field images of the thermally grown *as deposited* gold nano structures for this system and the corresponding selected area diffraction pattern. Due to non-wetting nature of gold on native oxide covered silicon substrate, isolated nanostructures of gold are formed on the surface (as shown in Fig. 5(a)). The average size of these gold nanostructures $34 \pm 1.02$ nm has a coverage area of 24%. The coverage area remained the same throughout the temperature range (up to a temperature of 700°C) showing no loss of gold from the surface. Comparing this with MBE grown structures (i.e., 2 nm Au case grown in UHV conditions), where the average size for the nanostructure is $27 \pm 1.1$ nm. The nano structures were also found to be isolated (i.e. not-connected network) in case of native oxide at the interface [27]. While carrying out the real-time imaging during the annealing procedures on this system (5 nm Au/SiO$_2$/Si (100)), no appreciable change in morphology was observed until 750°C. The native oxide at the interface hinders the inter-diffusion and hence the reaction rate for interacting with bulk silicon is reduced [27]. During MBE growth on clean surfaces, at relatively lower temperatures, the Au – Si interface allows inter-diffusion and in-plane diffusion in the system. When the specimen (system with oxide layer) was annealed at 800°C, agglomeration was observed at some locations. At 850°C, square /rectangular structures having average size $110 \pm 1.3$ nm were found along with the un-reacted gold particles (Fig. 5(c)). Contrast between the rectangular structure and some un-reacted gold particles can be clearly seen. It indicates the formation of some Au-Si compositions. Selected area diffraction shows the single crystalline silicon background and some weak polycrystalline rings of Au (Fig. 5(d)). Desorption of gold from the rectangular gold silicide structure can also be seen at one corner in Figure 5(c).

Similar structure is shown with a high resolution STEM in Figure 6. STEM images were used to determine the nature of these rectangular structures. The STEM-high angle annular dark field (HAADF) image shown in Figure 6(a) depicts a rectangular gold silicide structure. This shows the expected Z-contrast, so that the Au appears brighter than the Si substrate. Interestingly, at one corner we observe a decrease in the contrast value where a rectangular hole-like structure in larger island is observed. This suggests loss of gold atoms at this location. To analyze the local distribution of Au in these structures, STEM- X-ray energy dispersive spectrometry (XEDS) elemental mapping has been carried out for the whole structure shown in Figure 6(a). Figure 6(b) shows the Au elemental distribution and the Figure 6(c) depicts the Si elemental distribution. These maps not only confirm that the island consists of Au, but also that Au-depletion takes place at the island edges. The apparent Au-depletion in Figure 6, can be attributed to the melting induced desorption of gold from these silicide island structures. Figure 7 (a) depicts a STEM – HAADF image of one gold silicide structure which developed hole like structures and (b) is a TEM bright field image of the same structure shown in figure 7(a). One of the main issues in this study is to understand where and how the gold silicide desorbs and forms a pit like structure. As reported by Wang et al [10], the pits can be used to grow different kind of structures. Figure 7(c) depicts a high resolution lattice image taken at 300 kV using a Cs corrected microscope. It is to be noted that we could not succeed in taking lattice image from the same structure using 200 keV electrons (Figures 1 – 5 are carried out with 200 keV systems) as the structure appeared to be very thick. This demonstrates the importance of high energy electron microscope. The lattice spacing values suggest the presence of decomposed (or un-reacted) gold (with a lattice spacing of 0.235 nm) and a possible meta-stable phase of gold silicide (a value of 0.261 nm corresponding to $Au_5Si_2$(123) plane). Although lattice image from the center of the hole like structure is not evident, fast Fourier transform of area at the center showed some crystalline nature. More detailed studies are necessary to understand the evaluation of the hole like structures.

## 4. Conclusions

We showed that it is possible to grow symmetric (4-fold) gold silicide nano structures when grown under ultra high vacuum condition and on clean Si(100) surfaces and annealed *ex-situ* in a heating stage in TEM even at relatively low temperatures. Real time high temperature studies showed evaluation of these structures and the area coverage value and the aspect ratio of these structures were found to increase at higher temperatures. For system with oxide at interface, no symmetric structures were found until $\approx 750°$ C. At these higher temperatures, due to desorption of native oxide present at the interface, rectangular structures were formed. We observed desorption of gold and /or gold silicide structures and formation of symmetric hole like structures in some cases.

**Acknowledgements:** P V Satyam would like to thank the Department of Atomic Energy, Government of India for 11$^{th}$ plan project No. 11-R&D-IOP-5.09-0100 and the University of Bremen, Germany for his sabbatical visiting program.

**References:**

[1]   Cheung C L, Kurtz A, Park H and Lieber C M 2002 *J. Phys. Chem*. B **106** 2429

[2]   Helveg S, Lopez-Cartes C, Sehested J, Hansen P L, Clausen B S, Rostrup- Nielsen J R, Abild-Pedersen F and Norskov J K 2004 *Nature* **427** 426

[3]   Dick K A, Deppert K, Karlsson L S, Wallenberg L R, Samuelson L and Seifert W 2005 *Adv. Funct. Mater.* **15** 1603

[4]   Hannon J B, Kodambaka S, Ross F M and Tromp R M 2006 *Nature* **440** 69

[5]   Moyen E, Macé M, Agnus G, Fleurence A, Maroutian T, Houzé F, Stupakiewicz A, Masson L, Bartenlian B, Wulfhekel W, Beauvillain P and Hanbücken M 2009 *Appl. Phys. Lett.* **94** 233101

[6]   Bezryadin A, Dekker C and Schmid G 1997 *Appl. Phys. Lett.* **71** 1273

[7]   Morita T and Lindsay S 2007 *J. Am. Chem. Soc.* **129** 7262

[8]   Mativetsky J M, Burke S A, Fostner S and Grutter P 2007 *Small* **3** 818

[9]   Li B, Luo Z Q, Shi L, Zhou J P, Rabenberg L, Ho P S, Allen R A and Cresswell M W 2009 *Nanotechnology* **20** 085304

[10] Wang H, Zhang Z, Wong L M, Wang S, Wei Z, Li G P, Xing G, Guo D, Wang D and Wu T 2010 *ACS Nano* **4** 2901

[11] Whitesides G M and Grzybowski B 2002 *Science* **295** 2418-2421

[12] Hiraki A, Nicolet M A and Mayer J W 1971 *Appl. Phys. Lett.* **18** 178

[13] Narusawa T, Komiya S and Hiraki A 1973 *Appl. Phys. Lett.* **22** 389

[14] Hiraki A, Lugujjo E and Mayer J W 1972 *J. Appl. Phys*. **43** 3643

[15] Hiraki A, Shuto K, Kim S, Kammura W and Iwami M 1977 *Appl. Phys. Lett.* **31** 611

[16] Iwami M, Kim S, Kammura W and Hiraki A 1980 *Japan. J. Appl. Phy* **19** Suppl. 1 489

[17] Mayer J W and Tu K N 1974 *J. Vac. Sci. Technol.* **11** 86

[18]  Green A K and Bauer E 1974 *J. Appl. Phys.* **47** 1286

[19]  Le Lay G 1983 *Sur.Sci.* **132** 169

[20]  Baumann F H and Schroeter W 1988 *Philos. Mag. Lett.* **57** 75

[21]  Tersoff J and Tromp R M 1993 *Phys. Rev. Lett.* **70** 2782

[22]  Brongersma S H et al. 1998 *Phys. Rev. Letts.* **80** 3795–3798

[23]  Sekar K, Kuri G, Satyam P V, Sundaravel B, Mahapatra D P, Dev B N 1995 *Surf Sci.* **339** 96

[24]  Sekar K, Kuri G, Satyam P V, Sundaravel B, Mahapatra D P, Dev B N 1995 *Phys Rev.* B **51** 14330

[25]  Mundschau M, Bauer E, Telieps W 1989 *Surface science* **213** 381-392

[26]  Rout B, Sundaravel B, Das A K, Ghose S K, Sekar K, Mahapatra D P 2000 *J. Vac. Sci.Technol.* B **18** 1847

[27]  Bhatta U M, Dash J K, Roy A, Rath A, Satyam P V 2009 *J. Phys. Condens. Matter* **21** 205403; Bhatta U M, Dash J K, Rath A, Satyam P V 2009 *Applied Surface Science* **256** 567

[28]  Bhatta U M, Rath A, Dash J K, Ghatak J, Yi-Feng L, Liu C P, and Satyam P V, 2009 *Nanotechnology* **20** 465601

[29]  Kopatzki E, Giinther S, Nichtl-Peche W and Behm R J 1993 *Surface Science* **284** 154-166

[30]  Frank S, Wedler H, and Behm R J 2002 *Phys Rev.* B **66** 155435

[31] Vasisht Sanjeev and Shirokoff John 2010 *Applied Surface Science* **256** 4915–4923

[32]  Goswami D K, Satpati B, Satyam P V and Dev B N 2003 *Curr. Sci*. **84** 903

[33]  Ijima Sumio and Ichihashi T 1986 *Phys. Rev. Lett*. **56** 616

[34]  Ressel B , Prince K C, Homma Y and Heun S 2003 *J. Appl. Phys.* **93** 3886

[35]  Suryanarayana C, Anantharaman T R 1974 *Mater. Sci. Eng.* **13** 73

[36]  Anantharaman T R, Luo H L, Klement W 1966 *Nature* (London) **210** 1040

[37]  Gaigher H L and Vander Berg N G 1980 *Thin Solid Films* **68** 373

[38]  Green A K and Bauer E 1976 *J. Appl. Phys.* **47** 1284

[39]  Tsaur B Y and Mayer J W 1981 *Philos. Mag.* A **43** 345

[40]  $Au_7Si$ (JCPDS 26-0723), $Au_5Si$ (JCPDS 26-0725), $Au_5Si2$ (JCPDS 36-0938), $Au_2Si$ (JCPDS 40-1140), $Au_3Si$ (JCPDS 24-0463)

[41]  Hultman L, Robertson A, Hentzell T G, Enstroem I, Psaras P A 1987 *J. Appl. Phys.* **62** 3647

[42]  McCarthy D N and Brown S A 2008 *J. Phys. Conference series* **100** 072007

**Figure captions:**

**Figure 1:** As deposited MBE sample showing (a) gold nanostructures with typical size of about 24 nm (b) SAD pattern showing reflections of both Si and Au and corresponding (c) high resolution image of one of the islands which is showing the d-spacing of Au(111).

**Figure 2:** Bright field (BF) transmission electron micrographs depicting real time morphological changes during the in-situ heating. (a), (b), (c), (d), (e), (f),(g), (h) and (i) show the morphology at 200ºC, 300ºC, 325ºC, 350ºC, 363ºC, 400ºC, 510ºC, 600ºC and 700ºC respectively. With increasing temperature growth of rectangular nanostructures can be seen.

**Figure 3:** Real-time SAD patterns taken at various temperatures during heating. (a), (b), (c), (d), (e) and (f), show the real-time diffraction pattern at 325ºC, 350ºC, 363ºC, 510ºC, 600ºC and 700ºC, respectively.

**Figure 4:** (a) Bright field image taken at room temperature after the sample was heated up to 700ºC (b) selected area diffraction showing the presence of mixed phase of silicide and (c) high resolution transmission electron micrograph of single nano rectangles.

**Figure 5:** (a) Showing *as-deposited* thermally grown nanostructures 5nmAu/SiO$_2$/Si(100) system, (b) corresponding SAD pattern showing the reflection of Au and Silicon,(c) BF transmission electron micrograph at 850 ºC, and (d) Corresponding selected area diffraction pattern

**Figure 6:** (a) STEM – HAADF image on one of gold silicide structure. This shows formation of a hole like structure: (b) the elemental mapping formed with Au x-ray signals, (c) with Si x-ray signals in the STEM-XEDS mode.

**Figure 7:** (a) STEM – HAADF image on one of larger gold silicide structure. This shows formation of many holes like structures inside (b) the TEM bright field image of same structure that was shown in (a) and (c) the lattice image taken from the area shown with circles in (b).

**Table 1:** Average length of the nanostructures, the aspect ratio, and the percentage of having rectangular/square shape at various temperatures is mentioned in the tables.

**Figure 1** Rath A, *et al.*

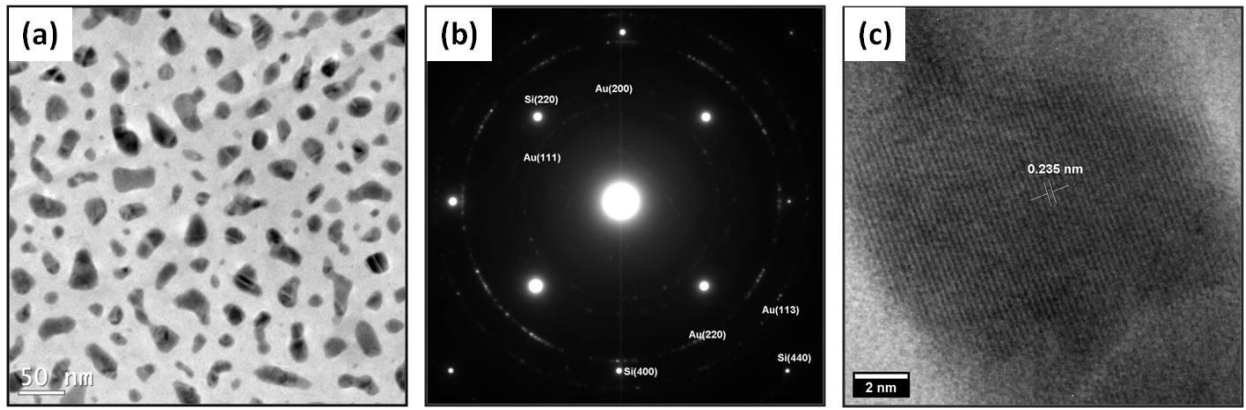

**Figure 2** Rath A, *et al.*

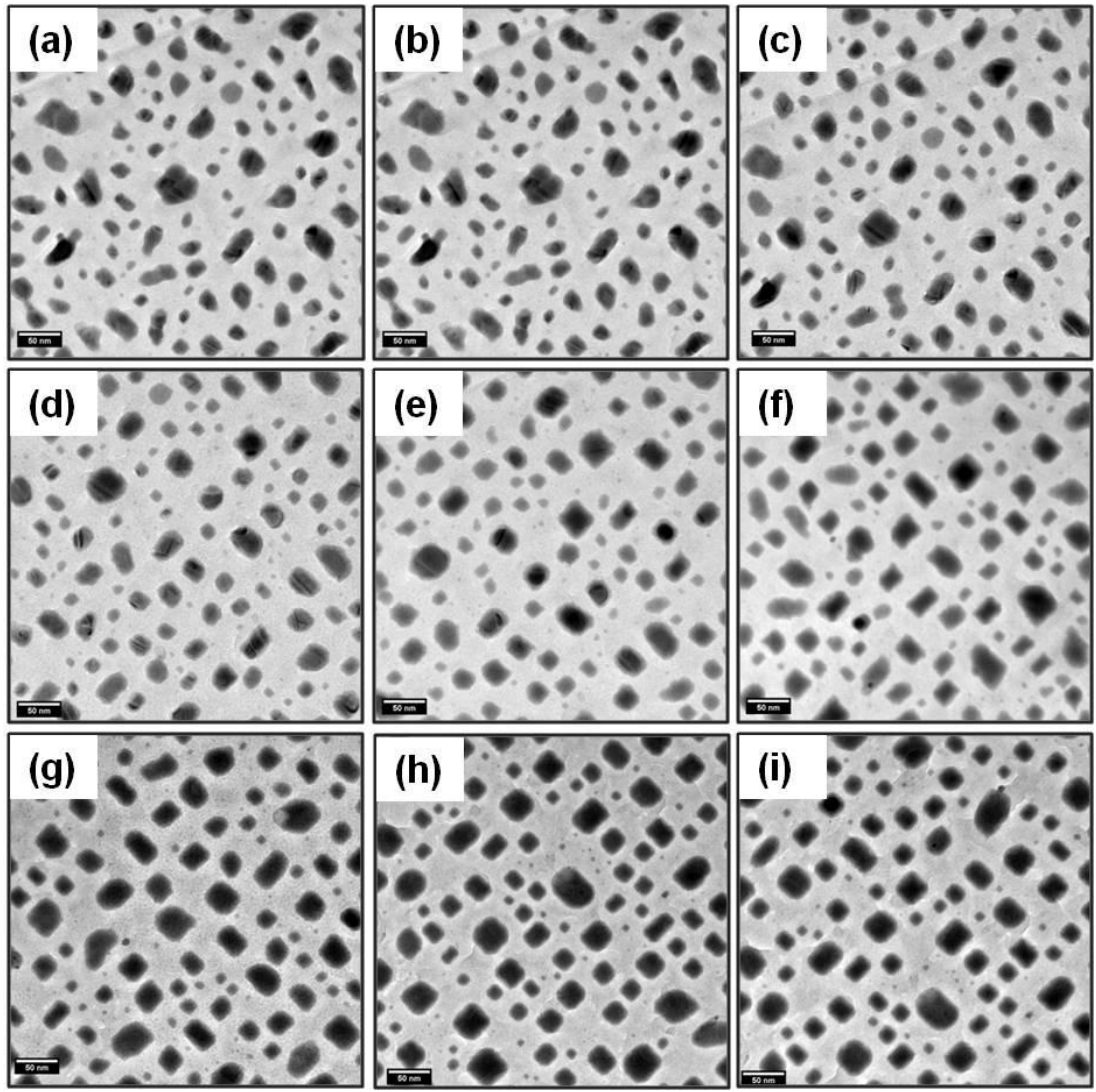

**Figure 3** Rath A, *et al.*

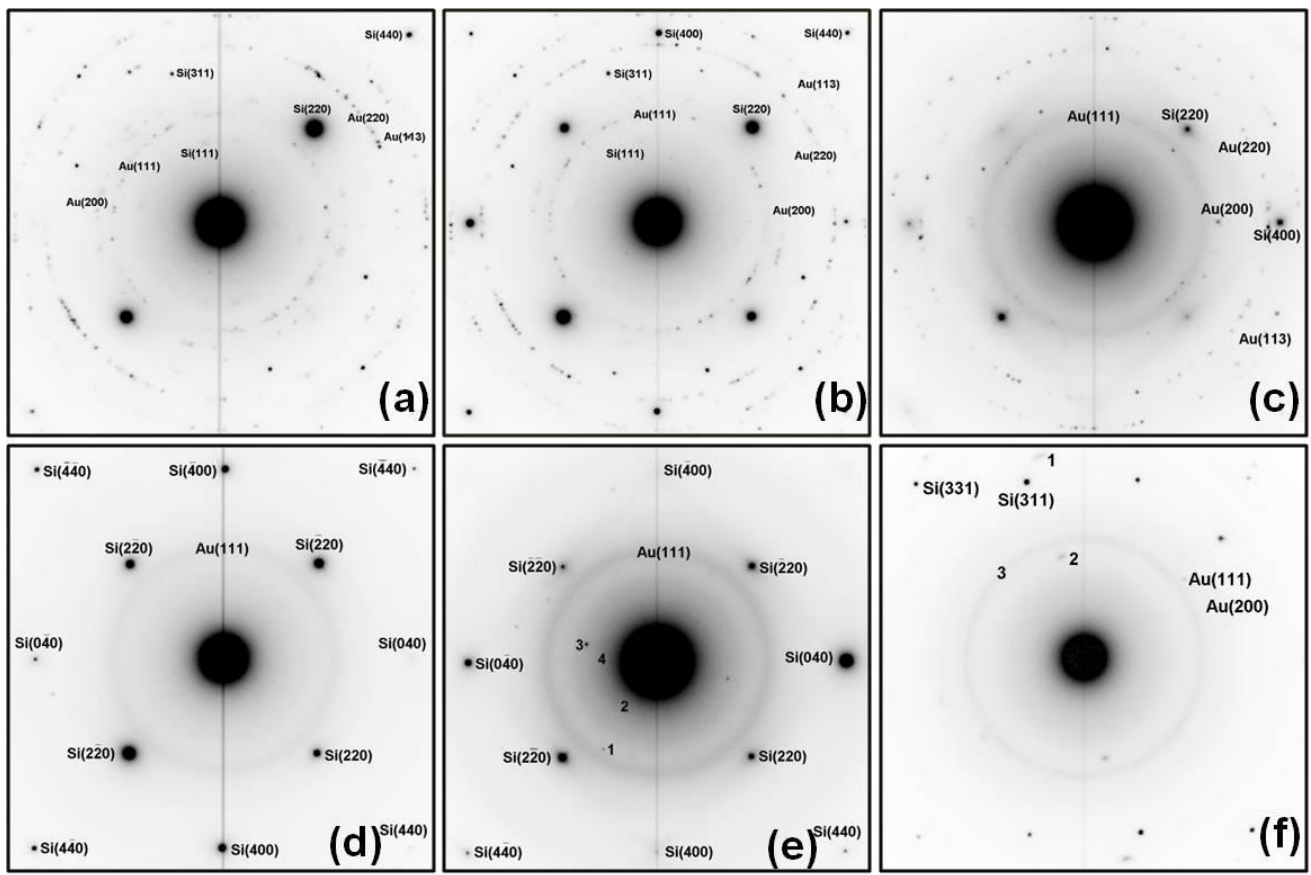

**Figure 4** Rath A, *et al.*

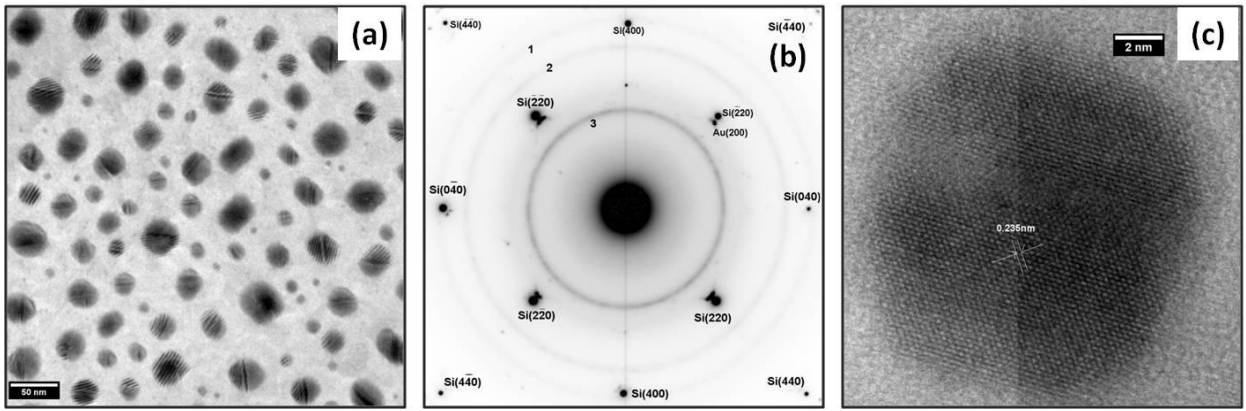

**Figure 5** Rath A, et al.,

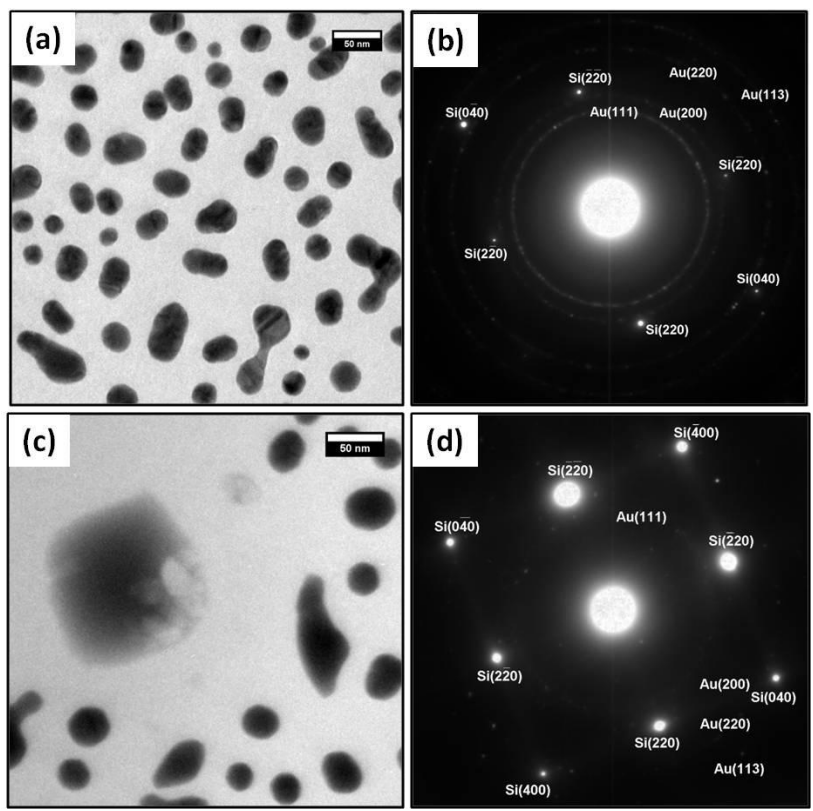

**Figure 6** Rath A, *et al.,*

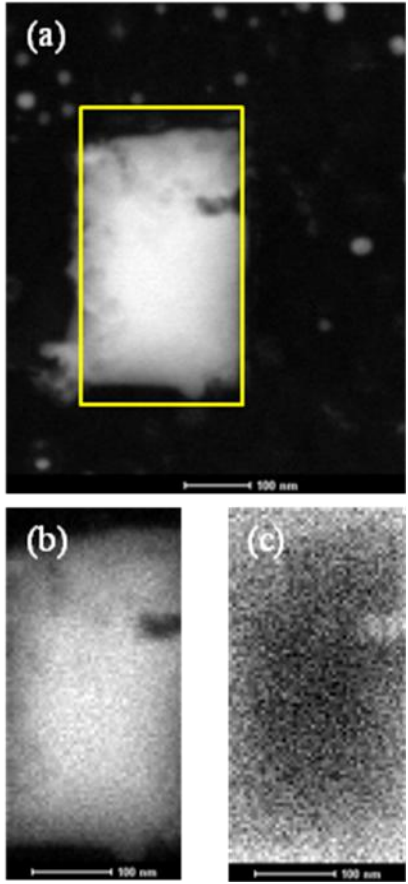

**Figure 7** Rath A, *et al.,*

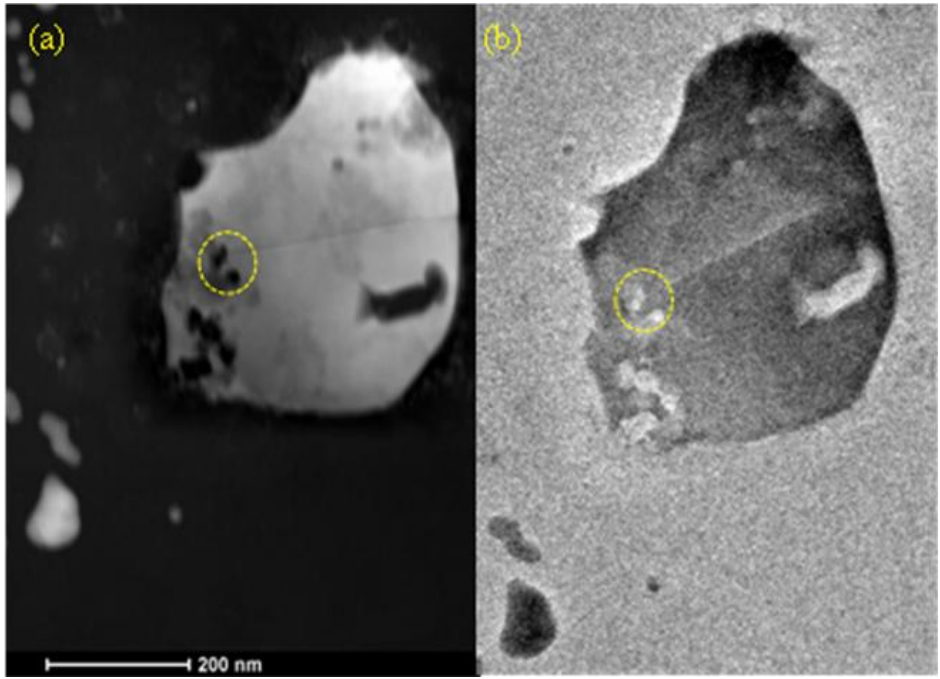

**Figure 7 (c) Rath** et al.,

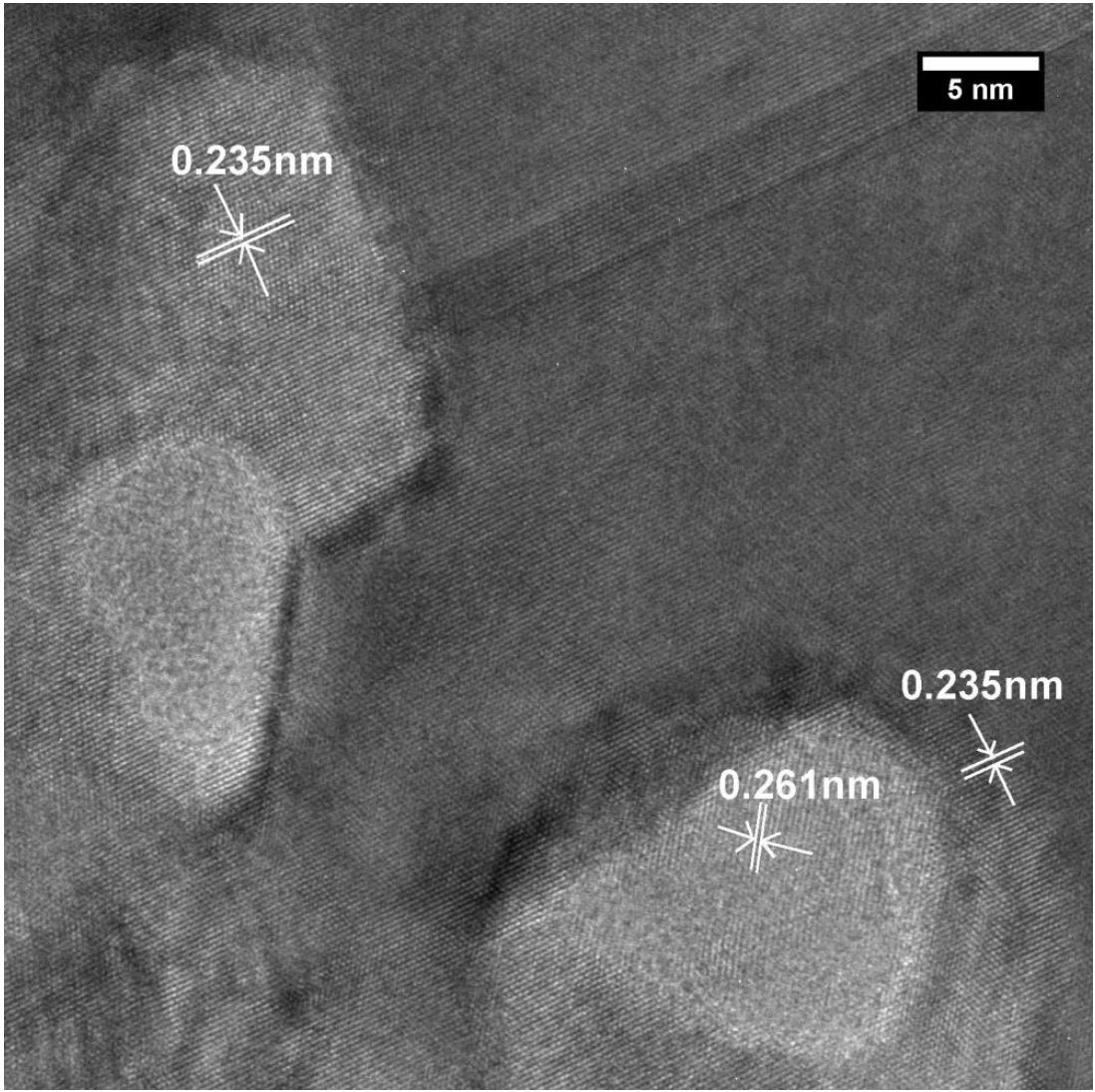

**Table 1**

| Temperature ($^0$C) | Average Length (nm) | Average aspect-ratio | Percentage of particle having rectangular/square shape |
|---|---|---|---|
| 325 | 19.5 ± 1.1 | 1.3 ± 0.02 | 36 |
| 350 | 20.2 ± 1.0 | 1.2 ± 0.01 | 45 |
| 363 | 24.3 ± 1.0 | 1.18 ± 0.01 | 58 |
| 400 | 26.7 ± 1.2 | 1.18 ± 0.02 | 69 |
| 510 | 25.1 ± 1.1 | 1.12 ± 0.01 | 80 |
| 600 | 24.9 ± 1.3 | 1.12 ± 0.01 | 81 |
| 700 | 25.9 ± 1.3 | 1.13 ± 0.02 | 82 |